\documentclass[10pt, conference, letterpaper]{IEEEtran}
\IEEEoverridecommandlockouts
\usepackage{cite}
\usepackage{amsmath,amssymb,amsfonts}
\usepackage{graphicx}
\usepackage{textcomp}
\usepackage{xcolor}

\usepackage{tabularx}
\usepackage{color, colortbl}
\usepackage[misc]{ifsym}
\usepackage{xcolor}
\definecolor{lime}{rgb}{0.88,2,10}

\newcommand*{\Resize}[2]{\resizebox{#1}{!}{$#2$}}%

\renewcommand{\baselinestretch}{.981}
\usepackage{subcaption}
\usepackage{wrapfig}
\usepackage[linesnumbered,ruled,vlined]{algorithm2e}
\usepackage{xpatch}
\usepackage{url}
\usepackage{multirow}
\usepackage{multicol}
\usepackage{wrapfig}

\usepackage[export]{adjustbox}
\usepackage{nomencl}
\usepackage{siunitx}
\usepackage{blindtext}
\usepackage[T1]{fontenc}
\usepackage[spaces,hyphens]{xurl}
\usepackage{float}

\usepackage{diagbox}

\SetAlFnt{\small}

\SetKwComment{Comment}{$\triangleright$\ }{}

\usepackage[utf8]{inputenc}
\usepackage{enumitem}

\SetKwInput{kwInit}{Initialize}

\usepackage{booktabs}

\def\BibTeX{{\rm B\kern-.05em{\sc i\kern-.025em b}\kern-.08em
    T\kern-.1667em\lower.7ex\hbox{E}\kern-.125emX}}

\usepackage[numbers,sort&compress]{natbib}
\usepackage[font={small}]{caption} 

\newcommand{\fref}[1]{Fig.~\ref{#1}}

\newcommand{\sref}[1]{Section~\ref{#1}}
\usepackage[misc]{ifsym}

\usepackage{tikz}
\newcommand*\circled[1]{\tikz[baseline=(char.base)]{
            \node[shape=circle,draw,inner sep=0.1pt] (char) {#1};}}
\newcounter{stepnum}

\usepackage[outline]{contour}
\contourlength{0.2pt}
\contournumber{15}

\newcommand\HUGE{\fontsize{25}{30}\selectfont}

\usepackage{fancyhdr}
\fancypagestyle{mystyle}{
    \chead{\large The 13th IEEE International Conference on Big Data (IEEE BigData 2025)}}

\begin{document}
\title{\HUGE{Secure and Privacy-Preserving Federated Learning for Next-Generation Underground Mine Safety} }


\author{\IEEEauthorblockN{
 Mohamed Elmahallawy\IEEEauthorrefmark{2}, Sanjay Madria\IEEEauthorrefmark{3}, Samuel Frimpong\IEEEauthorrefmark{4}}  
      \IEEEauthorblockA{%
 \IEEEauthorrefmark{2} School of Engineering and Applied Science, Washington State University, Richland, WA 99354, USA}
     \IEEEauthorblockA{%
 \IEEEauthorrefmark{3}Computer Science Department, Missouri University of Science and Technology, Rolla, MO 65401, USA}
   \IEEEauthorblockA{%
    \IEEEauthorrefmark{4}Explosive \& Mining
Engineering Department, Missouri University of Science and Technology, Rolla, MO 65401, USA}
 Emails:  {\{mohamed.elmahallawy, madrias@mst.edu, frimpong@mst.edu\}}}

\maketitle
\thispagestyle{mystyle}

\begin{abstract}
Underground mining operations depend on sensor networks to monitor critical parameters such as temperature, gas concentration, and miner movement, enabling timely hazard detection and safety decisions. However, transmitting raw sensor data to a centralized server for machine learning (ML) model training raises serious privacy and security concerns. Federated Learning (FL) offers a promising alternative by enabling decentralized model training without exposing sensitive local data. Yet, applying FL in underground mining presents unique challenges: (i) Adversaries may eavesdrop on shared model updates to launch model inversion or membership inference attacks, compromising data privacy and operational safety; (ii) Non-IID data distributions across mines and sensor noise can hinder model convergence. To address these issues, we propose \textit{FedMining}—a privacy-preserving FL framework tailored for underground mining. FedMining introduces two core innovations: (1) a \textit{Decentralized Functional Encryption (DFE)} scheme that keeps local models encrypted, thwarting unauthorized access and inference attacks; and (2) a \textit{balancing aggregation mechanism} to mitigate data heterogeneity and enhance convergence. Evaluations on real-world mining datasets demonstrate FedMining’s ability to safeguard privacy while maintaining high model accuracy and achieving rapid convergence with reduced communication and computation overhead. These advantages make FedMining both secure and practical for real-time underground safety monitoring.

\end{abstract}

\section{Introduction}


Underground mining presents inherently hazardous conditions due to its complex networks of tunnels and extreme environments where miners perform daily operations~\cite{r24}. Ensuring operational and personnel safety is a significant challenge, particularly given the harsh conditions and limited communication infrastructure in deep or confined spaces~\cite{goyal2022minerfinder}. As mining extends deeper underground, risks escalate—higher pressures can trigger catastrophic events such as wall collapses, pillar failures, and surface subsidence~\cite{debing2016current,zhang2016residual}, endangering both productivity and lives. To mitigate these risks, sensor networks are widely deployed to monitor critical parameters, including toxic gas levels, temperature, air quality, and structural stability~\cite{rahman2024cav,yadav2025predicting}. Additionally, these sensors facilitate accurate miner tracking {\em without GPS}, enabling rapid emergency response in scenarios such as fires or explosions~\cite{r48}.




%

In the era of AI technology, transmitting sensor data from various mines to a central server for training a deep machine learning (ML) model can enable real-time accurate detection of abnormal conditions, environmental hazards, and equipment malfunctions that impact miners' safety \cite{hyder2019artificial}. However, this approach raises serious concerns, including privacy breaches, data security risks, regulatory compliance challenges, and the potential misuse of sensitive information, making it a critical issue for mining operations. One promising solution is the adoption of federated learning (FL) \cite{mcmahan2017communication}, which allows multiple mine safety monitoring stations (MSMSs) to collaboratively train a global ML model in a decentralized manner while maintaining data privacy. In FL, each MSMS uses its own private data to train an ML model and shares only model weights or gradients, rather than the raw data, with an aggregator entity ($\mathcal{AE}$). The $\mathcal{AE}$ combines the received local models to create a global ML model, which is then distributed back to all MSMSs for further training. Iteratively, this process continues until the global model reaches convergence.

Although FL appears ``safe'' by sharing only model weights or gradients, it remains vulnerable to significant privacy and security threats. In FL, clients (mines) {\em wirelessly} transmit ML models over potentially {\em insecure} communication channels, exposing the system to threats from both insider adversaries and outsider eavesdroppers. Insider threats may arise when certain mines collude with the $\mathcal{AE}$ to bias the global model in their favor or to extract private information from other mines via model inversion or membership inference attacks. Outsider attackers may compromise model integrity, launch replay attacks~\cite{cao2022flcert,ghimire2022recent,alazab2021federated}, or exploit local updates through model stealing~\cite{nasr2019comprehensive,sun2021information}.

On the other hand, the quality of data collected by mines can vary significantly due to harsh underground conditions. Factors such as extreme darkness, poor visibility, and unreliable communication can lead to noisy or low-quality sensor and camera data~\cite{jewel2024dis}. Moreover, the non-independent and identically distributed (non-IID) nature of data from geographically dispersed mines exacerbates training difficulties \cite{sazedur2025detecting}. This variability can hinder global model convergence, often requiring hundreds or thousands of communication rounds, or even causing training failure.

To address these challenges, we propose {FedMining}—a secure federated learning framework for hazard detection and safety monitoring in underground mining, designed to protect the privacy of each MSMS's data. FedMining offers a \textit{secure aggregation scheme} that transforms traditional functional encryption (FE) into a lightweight, decentralized functional encryption (DFE) approach, avoiding the high overheads associated with homomorphic encryption (HE)~\cite{ma2022privacy} and multi-party computation (MPC)~\cite{kanagavelu2020two}. Moreover, to tackle data heterogeneity across MSMSs, FedMining proposes a \textit{balancing aggregation scheme} that mitigates bias from non-IID data distributions, ensuring the global model performs robustly across varied hazard and safety scenarios.
In summary, this paper introduces the following contributions: 
\begin{itemize}[leftmargin=*]
    \item We propose  {FedMining}, the first FL framework tailored for underground mining hazard detection that simultaneously ensures miner safety and safeguards the privacy of MSMSs’ local models and data. Unlike prior work, FedMining achieves high convergence accuracy in just a few communication rounds, making it both secure and efficient for real-world deployment.

 \item FedMining introduces two key innovations: (i) a \textit{secure aggregation scheme} that adapts traditional FE for efficient use in FL, protecting MSMSs' local models against both insider and outsider attacks. This ensures that the $\mathcal{AE}$ can construct a global hazard detection model without accessing any individual MSMS’s parameters—even over insecure channels—thereby preserving data privacy. (ii) a \textit{balancing aggregation scheme} that mitigates the effects of data imbalance and quality variation by weighting local updates based on participation frequency, dataset size, and model staleness. This eliminates bias toward larger datasets and boosts model accuracy across hazard scenarios.


   
    \item We conduct extensive experiments on real-world underground mining datasets, including the DsLMF+ dataset~\cite{yang2023open}, as well as CIFAR-10 and CIFAR-100~\cite{CIFAR-10}, to evaluate the security and performance of \textit{FedMining} in detecting diverse hazard scenarios in underground mining environments. The results show that FedMining significantly reduces both communication and computation overhead, outperforming Paillier-based aggregation schemes by 27$\times$ and 14$\times$, respectively. Additionally, it achieves over 93\% accuracy across all DsLMF+ dataset's classes with rapid convergence in just a few rounds.
\end{itemize}


\section{Related Work}


\subsection{Non-security-based hazard prediction}
While the introduction of ML for hazard prediction in this domain is still in its nascent stages, some pioneering works have laid the groundwork for such endeavors. For instance, in \cite{wojtecki2022use}, researchers predicted the rockburst hazard in an active hard coal mine using various models including neural network (NN), decision tree (DT), random forest (RF), gradient boosting, and extreme gradient boosting (XGB). Among these models, the NN  and DT demonstrated effectiveness, successfully classifying approximately 80\% of the rockbursts on average. In another study \cite{wojtecki2023attempt}, the authors introduced a prediction method for strong tremors during longwall mining, employing the RF model for both training and testing. Their approach yielded an average accuracy of 91.2\% in predicting minority cases. 
Moreover, in \cite{dey2021hybrid}, researchers proposed a hybrid convolutional neural network and long short term memory (CNN-LSTM) prediction model aimed at enhancing the safety and productivity of underground mines by extracting spatial and temporal features from mine data effectively. 
Another hybrid model named UMAP-LSTM combines uniform manifold approximation and projection (UMAP) with LSTM is proposed in \cite{kumari2021umap} to predict the fire status of sealed-off areas in underground mines for early warnings to protect miners' lives by anticipating potential mine hazards.   

\subsection{Security-based hazard prediction}

The literature has introduced preliminary efforts to safeguard the privacy of transmitted data among mines. For instance, the authors of \cite{dey2021deep} have proposed a secure CNN  architecture for underground mines, which predicts the presence of unauthorized individuals during voice communication. In \cite{wang2023pli}, the authors introduced a new privacy level indicator (PLI) aimed at mitigating the risk of privacy breaches associated with releasing original event logs without protection during the public data stage. PLI evaluates the trade-off between privacy enhancement and utility loss. To validate the efficacy of their method, the authors applied it to the underground locomotive dispatching system of an intelligent underground mine, providing strategic insights for the security application of critical data.

To the best of our knowledge, no prior work has proposed an FL framework specifically tailored for hazard detection or safety monitoring in underground mining, while simultaneously addressing the privacy concerns inherent to FL in this domain. Although several secure aggregation methods exist that could be adapted to this problem, each presents distinct advantages and limitations, as outlined below:

\begin{enumerate}[leftmargin=*, label=\arabic*.]
    \item {\bf Differential Privacy (DP) \cite{wei2020federated}.} DP is a non-cryptographic technique in which clients locally perturb their private data and transmit only the ``randomized'' version to $\mathcal {AE}$. This ensures that both other clients and the $\mathcal {AE}$ cannot retrieve the original private information. However, this method may reduce the prediction accuracy of the global model because of the added random noise.

    \item {\bf Multi-Party Computation (MPC) \cite{guo2020v}.} MPC allows all clients to share one-time pads to mask the data sent to $\mathcal {AE}$. To decrypt the data, $\mathcal {AE}$ collects a specified number of client shares. While this approach prevents the accuracy loss seen in DP-based methods, it requires additional interactions between $\mathcal {AE}$ and clients, thereby increasing communication and computation overheads.

    \item {\bf Homomorphic Encryption (HE) \cite{liu2021privacy}} HE allows computations to be performed on encrypted data without needing to decrypt it. In this approach, each client encrypts its local model, and then the $\mathcal {AE}$ carry out arithmetic operations on these encrypted models. However, as MPC, additional interactions between the $\mathcal {AE}$ and clients are necessary to decrypt and recover the global model.
\end{enumerate}

While existing schemes offer varying levels of privacy preservation for clients' local models, they often do so at the cost of reduced global model accuracy or increased computational and communication overhead. To date, no approach successfully achieves lightweight overhead, fast convergence, and high model accuracy simultaneously. In this paper, we propose a decentralized, lightweight encryption scheme that ensures data privacy while significantly improving convergence speed and maintaining high classification accuracy across all classes.

\section{FedMining System Model, Threat Model, And Design Objectives}


\subsection{Federated Learning in Underground Mining Operations}
We consider a set $\mathcal{K}$ of MSMSs deployed across different locations, each operated independently by a different vendor (and potentially located in different countries). Each MSMS $k$ gathers a dataset $D_k=\{x_{k,j},y_{k,j}\}_{j=1}^{|D_k|}$, containing $|D_k|$ samples which are used to train an ML model for specific classification tasks, such as detecting miner presence, monitoring safety helmet usage, and emergency support
systems. Here, $x_{k,j}$ represents the $j$-th data point collected by any MSMS $k$ which can be non-IID as compared to other MSMSs, while $y_{k,j}$ denotes the corresponding label. All MSMSs within the network $\mathcal K$ collaborate on training a global ML model with the aim of minimizing the following loss function:
\begin{equation}
    \min_{\boldsymbol{w}\in\mathbb{R}^d}\{\mathcal{L}(\boldsymbol{w})\}, ~\text{with}~\mathcal{L}(\boldsymbol{w})\triangleq\frac{1}{|D|}\sum_{k\in\mathcal{K}}|D_k|\mathcal{L}_k(\boldsymbol{w})
\end{equation} 
Here, $\boldsymbol{w}$ denotes a vector representing the global model parameters/weights. We define $\mathcal{L}_k(\boldsymbol{w})$ as the loss function of MSMS $k$,which can be expressed as:
\begin{equation}
\mathcal{L}_k(\boldsymbol{w})\triangleq\frac{1}{|D_k|}\sum_{\{x_{k,j},y_{k,j}\}\in D_{k}} l_{k}(\boldsymbol{w};x_{k,j},y_{k,j})
\end{equation} 
where $l_{k}(\boldsymbol{w};x_{k,j},y_{k,j})$ is the  training loss for model $\boldsymbol{w}$ on the data point $\{x_{k,j},y_{k,j}\}$. 

\subsection{Threat Model}\label{Sec:Threat}

Our threat model encompasses both internal and external adversaries, which we define as follows:
\begin{itemize}[leftmargin=*]
   \item {\bf   Internal adversaries:} MSMSs and the $\mathcal{AE}$ are considered ``semi-honest'' participants, meaning they adhere to the FL protocol honestly but harbor a curiosity to learn or infer sensitive information about the private raw data of some or all other MSMSs' local models. Additionally, we consider the potential for collusion between the $\mathcal{AE}$ and a subset of MSMSs owned by different operators, aiming to train a biased global model favoring this subset or learning local models of other MSMSs.

  \item  {\bf External adversaries:}  An adversary may eavesdrop on the communication channels between MSMSs and the $\mathcal{AE}$ to steal transmitted information (i.e., local models). Additionally, the adversary could launch membership inference or model reverse attacks to obtain private information about the MSMSs' data, such as the amount of mineral resources, number of miners, etc. 
\end{itemize}


\subsection{Design Objectives}

\noindent{\bf $\mathcal{O}(1)$~Data and Model Confidentiality:} The scheme must ensure the confidentiality of both the MSMSs' local model parameters and their associated private data, preventing any information leakage to the $\mathcal{AE}$, other MSMSs, eavesdroppers, or adversaries.


    
\noindent{\bf $\mathcal{O}(2)$~Efficiency and Accuracy:} The scheme should maintain low communication and computational overhead during local model encryption and global model generation. Moreover, the global model must achieve high hazard detection accuracy across all target scenarios and classes, despite non-IID data distributions from geographically distributed MSMSs.


\section{ Methodology}\label{Sec:Proposed}

\subsection{FedMining Framework}
\begin{figure*}[!t]
\centering
\includegraphics[width=0.8\textwidth]{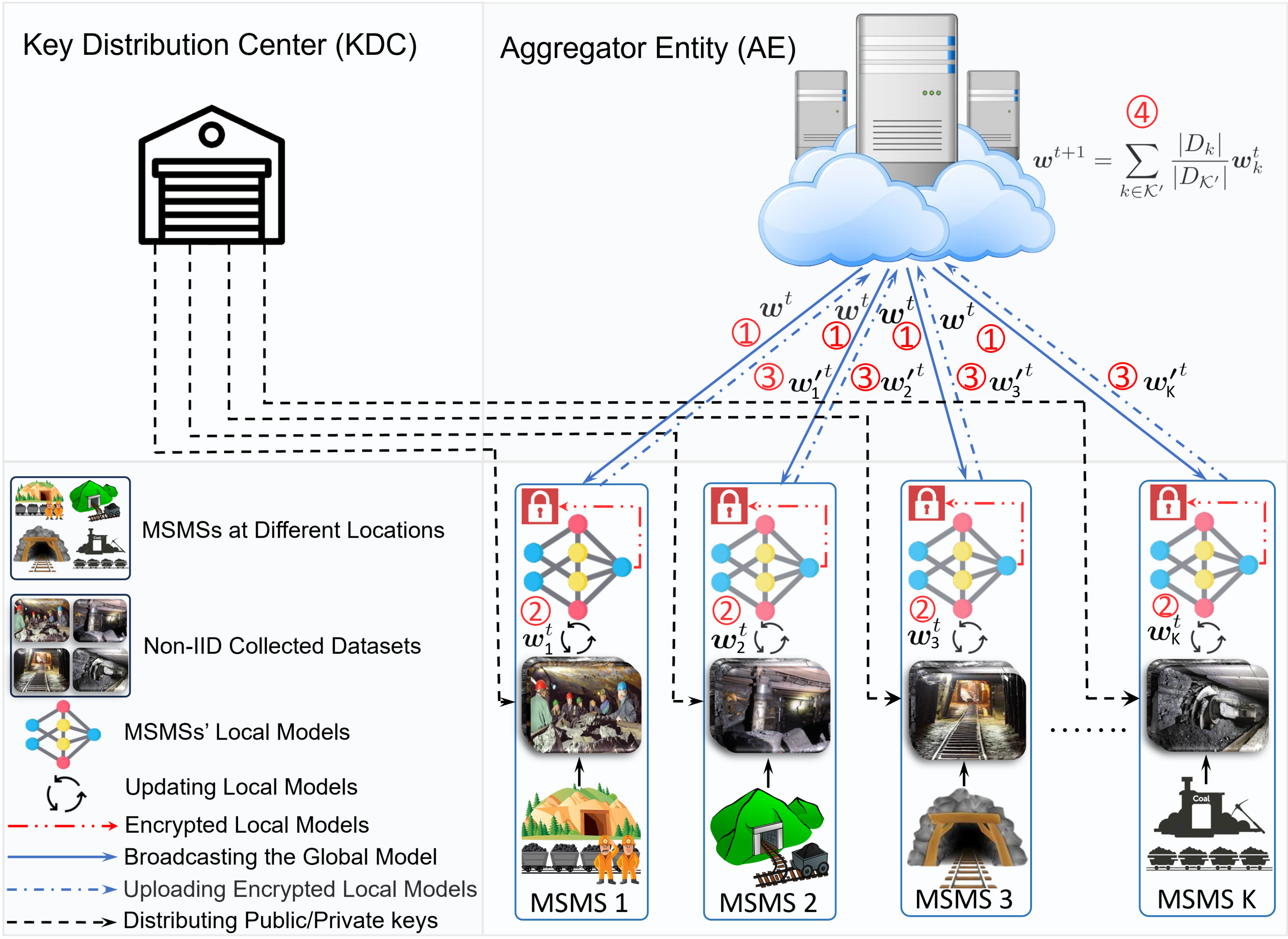}
\caption{An illustrative visualization of the FedMining framework.} \label{System_Model}
\end{figure*}

FedMining is a secure FL framework designed to train a global ML model for hazard detection in underground mining, while safeguarding the privacy of each MSMS's local model from potential attacks. This approach is unique in that it not only protects the privacy of MSMSs' local models and their associated data but also ensures rapid convergence of the FL process. Importantly, it effectively mitigates biases in the global model caused by MSMSs with larger datasets and adeptly manages the non-IID nature of the data. This guarantees high accuracy across {\em all target classes} without giving undue preference to any single MSMS's dataset. Moreover, FedMining's secure aggregation scheme demonstrates efficient security computations and communication overheads. The training process of FedMining involves four steps in each communication round $t$, as illustrated in \fref{System_Model},  which are described as follows:

\noindent{\bf Stage \circled{1}} The $\mathcal{AE}$ distributes the latest global model $\boldsymbol{w}^{t}$ to all available MSMSs that have agreed to engage in the current FL training round $t$. Note, at $t=1$, the model $\boldsymbol{w}^{1}$ is initialized randomly before dissemination.
    
\noindent{\bf Stage \circled{2}} Each available MSMS utilizes the received $\boldsymbol{w}^{t}$ to retrain its ML model using its private data using a local optimizer like stochastic gradient descent (SGD). This process yields an updated local model $\boldsymbol{w}_{k}^{t}$ through local epochs, denoted as $i=1,\dots,I$, following the iterative update as $\boldsymbol{w}_{k}^{t,i+1}=\boldsymbol{w}_{k}^{t,i}- \eta_{t} \nabla \mathcal{L}_{k} (\boldsymbol{w}_{k}^{t,i})$, where $\eta_{t}$ is the learning rate.

\noindent{\bf Stage \circled{3}} Each participating MSMS  then uses its private key, which is generated and distributed by the key distribution center (KDC), to encrypt its local model  $\boldsymbol{w}_{k}^{t}$ using our secure aggregation scheme (\sref{Sec:secagg}) to generate a secure version $\boldsymbol{w'}^{t}_{k}$, and then uploads $\boldsymbol{w'}^{t}_{k}$ to the $\mathcal{AE}$.
    
\noindent{\bf Stage \circled{4}} Once the $\mathcal{AE}$ collects all MSMSs' local models, it utilizes the aggregation key (see \sref{Sec:secagg}) to aggregate their weighted average as $ \boldsymbol{w}^{t+1}= \sum_{k\in \mathcal{K}'} \frac{|D_k|}{|D_{\mathcal{K}}|} \boldsymbol{w}_{k}^{t}$, where $\mathcal{K}$ denotes the set of all participating MSMSs with a total data size of $|D_{\mathcal{K}}|$. In \sref{Sec:balance}, we will explain our aggregation scheme for improved convergence accuracy across all classes, replacing the weighted average $\frac{|D_k|}{|D_{\mathcal{K}}|}$ by a new weighting $\alpha_{k}^{t}$.


Finally, the $\mathcal{AE}$ transmits the aggregated model,  $\boldsymbol{w}^{t+1}$, back to the next available MSMSs in the subsequent rounds similar to Stage \circled{1}. This iterative process continuous over rounds $T$ until the global model converges. 

\subsection{Secure Aggregation Scheme}\label{Sec:secagg}




Our secure aggregation scheme leverages inner-product functional encryption (IPFE), which restricts the decryptor (i.e., $\mathcal {AE}$ in our scenario) capabilities to computing only the inner product on encrypted data (i.e., MSMSs' local models). 
Specifically, we adapt the IPFE scheme proposed by Bishop et al. \cite{bishop2015function} to facilitate efficient privacy-preserving data aggregation in FedMining. While this IPFE scheme employs a master secret key ($\mathcal{MSK}$) to encrypt the data, our scenario involving various MSMSs necessitates unique secret keys for each one to maintain security. Therefore, we partition the  $\mathcal{MSK}$ into multiple distinct keys, assigning one to each MSMS. During encryption, each MSMS utilizes its designated secret key 
to generate a ciphertext. Subsequently, during decryption, the ciphertexts from various MSMSs are aggregated by the $\mathcal {AE}$ using its aggregation key to generate the global model in plaintext. This adaptation consists of three phases: initialization, encrypting MSMSs' local models, and secure model aggregation, as follows:

\begin{enumerate}[leftmargin=*,label={\bf \arabic*.}, leftmargin=0pt, labelwidth=*, labelsep=0.5em, align=left]
\item{\bf Initialization Phase.} The KDC generates public and secret parameters for each MSMS and the $\mathcal{AE}$, respectively, as outlined below:
\begin{enumerate}[leftmargin=*, label={(\alph*)}, leftmargin=0pt, labelwidth=*, labelsep=0.5em, align=left]
\textbf{ \item Setup}($1^\alpha$, $S$) $\rightarrow$ ($\mathcal{PP}$, $\mathcal{MSK}$): This procedure takes as input a security parameter $\alpha$ and a set $S \subseteq \mathbb{Z}_q$, and outputs the public parameters ($\mathcal{PP}$) and the  $\mathcal{MSK}$ as:
\begin{itemize}[leftmargin=*]
    \item First, it chooses two cyclic additive groups $\mathbb{G}_1$, $\mathbb{G}_2$, along with a cyclic multiplicative group $\mathbb{G}_T$, all of prime order $q$, and defines an asymmetric bilinear pairing function $e: \mathbb{G}_1 \times \mathbb{G}_2 \rightarrow \mathbb{G}_T$ that maps elements from $\mathbb{G}_1$ and $\mathbb{G}_2$ to an element in $\mathbb{G}_T$.
    \item  It then chooses randomly two generator elements $\mathrm{g}_1$ and $\mathrm{g}_2$,where $\mathrm{g}_1 \in \mathbb{G}_1$ and $\mathrm{g}_2 \in \mathbb{G}_2$.

    \item Next, it computes three invertible matrices over the finite field $\mathbb{Z}_q$, denoted as $\{\mathbf{B}$, $\mathbf{A_1}$, $\mathbf{A_2}\}$, each with dimensions $n \times n$, which are determined by the size of the local models of the MSMSs.


    \item Finally, the KDC generates the $\mathcal{PP}$ as $(\mathbb{G}_1, \mathbb{G}_2, \mathbb{G}_T, q, e,\mathrm{g}_1 , \mathrm{g}_2)$ and the  $\mathcal{MSK}$ as $(\mathbf{B}, \mathbf{A_1}, \mathbf{A_2})$, which is kept secret (never shared with any other MSMS).
\end{itemize}
\textbf{ \item  KeyGenMSMS}$(\mathcal {MSK}, k)\rightarrow (\boldsymbol{\lambda}_k,\mathcal{SK}_{k})$: The key generation procedure takes the master secret key $\mathcal{MSK}$ as input and produces a masking vector $\boldsymbol{\lambda}_k \in \mathbb{Z}_q^n$ for each $k$-th MSMS to obfuscate its local model parameters. It also generates a secret key $\mathcal{SK}_{k}$, which is used to encrypt each parameter as follows:

\begin{itemize}[leftmargin=*]
    \item The KDC assigns a unique $\boldsymbol{\lambda}_k$ to each $k$-th MSMS and computes the $\mathcal{SK}_{k}$ for each $k$-th MSMS as $
    \mathcal{SK}_{k} = ({{\mathbf{A_1}}^{\!\!\!\!-1}\mathbf{B}_{k}^{'}, \mathbf{A_2}^{\!\!\!\!-1}\mathbf{B}_{k}^{''}})$, where $\mathbf{B}_{k}^{'}$ and $\mathbf{B}_{k}^{''}$ are two invertible  matrices with $\mathbf{B}_{k}^{'}+\mathbf{B}_{k}^{''}=\mathbf{B}^{-1}$.
    
    
    \item Next, before distributing the secret key $\mathcal{SK}_{k}$ 
 and the parameter $\boldsymbol{\lambda}_k$ to each $k$-th MSMS, the KDC encrypts them using the public key and subsequently sends them to the respective $k$-th MSMS.
    
    \item Once each MSMS receives its encrypted secret keys,  it utilizes its private key to decrypt them.
    
\end{itemize} 
\textbf{ \item KeyGen}$\boldsymbol{\mathcal{AE}}(\mathcal {MSK})\rightarrow (\boldsymbol{\lambda},\mathcal{AK}_{\mathcal{AE}})$: The key generation procedure utilizes $\mathcal{MSK}$ as input to generate the $\mathcal{AE}$'s aggregation key, which is used to aggregate the received encrypted model parameters. Additionally, $\boldsymbol{\lambda}$, an n-dimensional vector, is generated to unmask the aggregated parameters. These steps are outlined as:

\begin{itemize}[leftmargin=*]
   \item  First, the KDC computes the $\mathcal{AK}_\mathcal{AE}$ and $\boldsymbol{\lambda}$ as $
       \mathcal{AK}_{\mathcal{AE}} = (\mathbf{B}\mathbf{A}_{1}, \mathbf{B}\mathbf{A}_{2})$ and $
       \boldsymbol{\lambda}=\sum_{i=1}^{K}\boldsymbol{\lambda}_{i}$.
   \item  The ${\mathcal{AE}}$ notifies the KDC of the available MSMSs, triggering the KDC to provide the corresponding $\boldsymbol{\lambda}$ in response. 
   \item Then, the KDC distributes the ${\mathcal{AE}}$'s keys, $\mathcal{AK}_{\mathcal{AE}}$ and $\boldsymbol{\lambda}$, after encrypting them using ${\mathcal{AE}}$'s public key to the ${\mathcal{AE}}$.
   \item Once the ${\mathcal{AE}}$ receives its encrypted keys, it utilizes its private key to decrypt them.
\end{itemize} 

\end{enumerate}

\item{\bf Encrypting MSMS's Local Models Phase.} 
Once each MSMS receives its public and private keys from the KDC, it performs the following two steps:
\begin{enumerate}[leftmargin=*, label={(\alph*)}, leftmargin=0pt, labelwidth=*, labelsep=0.5em, align=left]
\item  Each $k$-th MSMS trains a local ML model using its private data until convergence is achieved.

\item Then, each $k$-th MSMS masks and encrypts the generated model parameters using the following encryption procedure:

$\textbf{Encrypt}(\boldsymbol{\lambda}_k,\mathcal{SK}_{k}, \boldsymbol{w}_k^t)\rightarrow \boldsymbol{w'}^{t}_{k}$:
The encryption procedure takes $\boldsymbol{\lambda}_k$, $\mathcal{SK}_{k}$, and $\boldsymbol{w}_k^t$ for each $k$-th MSMS, and  generates  $\boldsymbol{w'}^{t}_{k}$ as an output which represents the ciphertext of $\boldsymbol{w}_k^t$. Specifically, it first adds the mask as
\begin{equation}\label{eqn:mask}
\widehat{\boldsymbol{w}_k^t}=\mathcal{S}_k\boldsymbol{w}_k^t+\boldsymbol{\lambda}_k
\end{equation}
where $\mathcal{S}_k=\frac{|D_k|}{|D_{\mathcal{K}}|}$ is the ratio between the data size MSMS $k$ and the overall data sizes among all MSMSs. Then, it encrypts these masked weights, $\widehat{\boldsymbol{w}_k^t}$, as 
\begin{equation}
    \boldsymbol{w'}^{t}_{k}=(\widehat{\boldsymbol{w}_k^t}\mathbf{A}_{1}^{\!\!-1} \mathbf{B}'_{k}\mathrm{g}_2, \widehat{\boldsymbol{w}_k^t}\mathbf{A}_{2}^{\!\!-1}\mathbf{B}''_{k}\mathrm{g}_2)
\end{equation}
\end{enumerate}

\item{\bf Secure Model Aggregation Phase.} 
For each FL round $t$, once the $\mathcal{AE}$ receives the encrypted local models from all MSMS, the $\mathcal{AE}$ will securely aggregate them as follows\footnote{The $\mathcal{AE}$ can only compute the aggregation of MSMSs' local models to generate the global model in plaintext, inferring nothing about  each individual MSMS's local model.}:

$\textbf{Decrypt}(\mathcal{AK}_{\mathcal{AE}},\boldsymbol{\lambda},\{\boldsymbol{w'}^{t}_{1},\dots,\boldsymbol{w'}^{t}_{K}\})\rightarrow \boldsymbol{w}^{t}$: The decryption procedure takes $\mathcal{AK}_{\mathcal{AE}}$, $\boldsymbol{\lambda}$, and the ciphertext of the MSMSs' models $\{\boldsymbol{w'}^{t}_{1},\dots,\boldsymbol{w'}^{t}_{K}\}$ as input, and generates $\boldsymbol{w}=\sum_{k\in\mathcal K} {\boldsymbol{w}_{k}}$, which can be calculated as:
\begin{align}
 \Resize{0.5cm}{\boldsymbol{w'}^{t}}&=\Resize{6cm}{{\sum_{k\in\mathcal{K}}\boldsymbol{w'}^{t}_{k}=}\biggl({\underbrace{\sum_{k\in\mathcal{K}}\widehat{\boldsymbol{w}_k^t}\mathbf{A}_{1}^{\!\!-1} \mathbf{B}'_{k}\mathrm{g}_2}_{\boldsymbol{w'}^{t}({1})},\underbrace{ \sum_{k\in\mathcal{K}}\widehat{\boldsymbol{w}_k^t}\mathbf{A}_{2}^{\!\!-1}\mathbf{B}''_{k}\mathrm{g}_2}_{\boldsymbol{w'}^{t}(2)}}\biggl) }
\end{align}
By applying the bilinear pairing to the above equation, we obtain the following result:
\begin{align}
 e(\mathbf{B}\mathbf{A}_{1}\mathrm{g}_1,\boldsymbol{w'}^{t}({1}))=e({\mathrm{g}_1,\mathrm{g}_2})^{\bigl(\sum_{k\in\mathcal{K}}\widehat{\boldsymbol{w}_k^t} \mathbf{B}'_{k}\bigl)\mathbf{B}}\label{eqn:11}\\
    e(\mathbf{B}\mathbf{A}_{2}\mathrm{g}_1,\boldsymbol{w'}^{t}({2}))=e({\mathrm{g}_1,\mathrm{g}_2})^{\bigl(\sum_{k\in\mathcal{K}}\widehat{\boldsymbol{w}_k^t} \mathbf{B}''_{k}\bigl)\mathbf{B}}\label{eqn:12}
\end{align}
By multiplying \eqref{eqn:11} and \eqref{eqn:12}, we get:
\begin{align}\label{eqn:mult}
 \Resize{7.5cm}{{e({\mathrm{g}_1,\mathrm{g}_2})^{\bigl(\sum_{k\in\mathcal{K}}\widehat{ \boldsymbol{w}_k^t} (\mathbf{B}'_{k}+\mathbf{B}''_{k})\bigl)\mathbf{B}}
=e({\mathrm{g}_1,\mathrm{g}_2})^{\bigl(\sum_{k\in\mathcal{K}}\widehat{\ \boldsymbol{w}_k^t}\bigl)}}}
\end{align}
  Given that $\widehat{\boldsymbol{w}_k^t}=\mathcal{S}_k\boldsymbol{w}_k^t+\boldsymbol{\lambda}_k$, then equation \eqref{eqn:mult} can be simplified as:
\begin{align}    
{e({\mathrm{g}_1,\mathrm{g}_2})^{\bigl({\sum_{k\in\mathcal{K}}{\boldsymbol \lambda}_k}+\sum_{k\in\mathcal{K}}\mathcal{S}_k{\boldsymbol{w}_k^t}\bigl)}=e({\mathrm{g}_1,\mathrm{g}_2})^{\bigl(\boldsymbol \lambda+\boldsymbol{w}^t\bigl)}}\label{eqn:13}
\end{align}
By utilizing $\mathrm{g}_1$ and $\mathrm{g}_2$, and employing the bilinear function, we can compute the following:
\begin{align}\label{eqn:14}
e(\boldsymbol{\lambda}{\mathrm{g}_1,\mathrm{g}_2})=e( {\mathrm{g}_1,\mathrm{g}_2})^{ \boldsymbol{\lambda} }
\end{align}
By dividing \eqref{eqn:13} and \eqref{eqn:14}, we obtain:
\begin{align}
e({\mathrm{g}_1,\mathrm{g}_2})^{\boldsymbol{\lambda}+{\boldsymbol{w}^t}-\boldsymbol{\lambda}}=e({\mathrm{g}_1,\mathrm{g}_2})^{\boldsymbol{w}^t}.
\end{align}
The $\mathcal{AE}$ then employs a discrete logarithm method, such as {\em Pollard's rho algorithm} \cite{teske1998speeding} or {\em the baby-step giant-step algorithm} \cite{shoup1997lower}, to compute the global model, $\boldsymbol{w}^t=\sum_{k\in \mathcal{N}}\mathcal{S}_{k}\boldsymbol{w}_{k}^{t}$, from $e({\mathrm{g}_1,\mathrm{g}_2})^{\boldsymbol{w}^t}$, which represents the weighted aggregate model parameter, considering that the parameters of $\boldsymbol{w}^t$ are small in their size.

\end{enumerate} 
\subsection{Balancing Aggregation Scheme}\label{Sec:balance}

Although our secure aggregation scheme (\sref{Sec:secagg}) weights the global model $\boldsymbol{w}^t$ based on the data size collected by each MSMS, this alone does not guarantee convergence or ensure satisfactory classification accuracy across all target classes. The challenge arises because, in each round $t$, some MSMSs participate in generating the global model, while others may be unavailable or unwilling to participate, though they might rejoin in future rounds. Additionally, participating MSMSs may contribute lower-quality models, which could negatively impact the global model’s convergence. We refer to these MSMSs as ``unreliable MSMSs'', categorized as follows:
\begin{enumerate}[leftmargin=*, label={\bf \arabic*.}]
    \item {\bf MSMSs with stale models.} These MSMSs participate inconsistently in FL rounds, leading them to train their local models using {\em outdated versions of the global model}. This inconsistency in participation will negatively impact global model convergence.

    \item {\bf MSMSs with low-quality data:} Training on such data results in local models of low accuracy. Despite the lower quality of their data or models, these MSMSs cannot be neglected or omitted from participation in the global model aggregation. They contribute to improving global model diversity, especially if they capture data distributions with different hazard scenarios/classes that other MSMSs fail to capture. For instance, a low-end camera may not produce high-resolution images but could capture infrequent classes of images that other MSMSs do not have \cite{jewel2024dis}.
\end{enumerate}
Excluding these types of MSMSs would undermine the heterogeneous FL setting among all MSMSs. Moreover, favoring MSMSs with fresh models or larger data sizes over those with stale models {\em could introduce a bias towards MSMSs with greater data sizes and more frequent participation.}


To address this issue, we propose this component that ensures equitable contribution from all MSMSs to the global model, minimizing bias without affecting its accuracy and facilitating rapid convergence across rounds. Specifically, for each MSMS $k$ willing to participate in the current FL round, after receiving the global model $\boldsymbol{w}^t$ from the $\mathcal{AE}$ along with statistical information about the data sizes of other participating MSMSs, each MSMS updates its local model using $\boldsymbol{w}^t$ and its private data. Instead of solely relying on $\mathcal{S}_k$ to mask its local model before generating the ciphertext as described in \eqref{eqn:mask}, each MSMS now incorporates {\em a new weighting factor}. This factor accounts for its participation frequency, the freshness of its local model, and its data size ratio as follows:
\begin{align}\label{eq:weighting}
\begin{split}
\alpha_{{k}}^{t}=\frac{\beta_{{k}}^{t}}{\sum_{{k}\in \mathcal{K}'}\beta_{{k}}^{t}},\quad{\beta_{{k}}^{t}}=\frac{f_{{k}}^{t -1}}{\sum_{{k}\in \mathcal{K}'} f_{{k}}^{t-1}}\times \mathcal{S}'_k
\end{split}
\end{align}
where $\alpha_{{k}}^{t}$ is a weighting factor calculated by each MSMS ${k}$, considering the staleness of its local model and its data size. $f_k^{t-1}$ is the participation frequency of MSMS $k$ up to round $t-1$ with $t-\gamma\leq f_{k}^{t-1}\leq t-1$, where $\gamma$ is a tolerance factor controlling the acceptable level of staleness. Here, $\mathcal{S}'_k=\frac{|D_k|}{|D_{\mathcal{K}'}|}$ denotes  the ratio between the data size of MSMS $k$ and the overall data size of the {\em currently} participating MSMSs $\mathcal{K}'$.

Please note, we can adjust the tolerance factor $\gamma$ to achieve a more balanced aggregation while addressing model staleness. For example, setting $\gamma=t$ allows for the inclusion of all MSMSs in the aggregation process, even those with the lowest participation frequencies. However, their contributions are scaled down by assigning them lower weights, which mitigates their negative impact on the global model.


Once each MSMS computes its $\alpha_{k}^{t}$ in round $t$, it replaces $\mathcal{S}_k$ in \eqref{eqn:mask} with $\alpha_{k}^{t}$. Consequently, \eqref{eqn:mask} can be updated as follows:
\begin{equation} 
\widehat{\boldsymbol{w}_k^t}=\alpha_{{k}}^{t}\boldsymbol{w}_k^t+\boldsymbol{\lambda}_k
\end{equation}
After that, the rest of the encryption and decryption process remains the same until the $\mathcal {AE}$ calculates the global model in plaintext as follows:
\begin{equation}
    \boldsymbol{w}^t=\sum_{k\in \mathcal{K}}\alpha_{k}^{t}\boldsymbol{w}_{k}^{t}
\end{equation}

\section{Performance Evaluation}


\subsection{Experimental setup}

{\bf Dataset and FL setting.} 
We utilize the DsLMF+  underground mining dataset \cite{yang2023open} for both training and validation tasks. This dataset comprises 138,004 annotated images across six categories: mine personnel, hydraulic support guard plate, large coal, towline, miners' behavior, and mine safety helmet. 

We also consider a cross-silo FL setting with 12 MSMSs at different locations, where each MSMS trains the YOLOv7 model architecture with 37,196,556 parameters to classify the six classes. Additionally, we utilize the CIFAR-10 and CIFAR-100 datasets \cite{CIFAR-10}, which are commonly used in evaluating state-of-the-art (SOTA) methods in the literature, to evaluate and compare our approach. For these datasets, each MSMS trains a two-layered CNN model followed by three fully connected layers with 7,759,521 trainable parameters. Our training process encompasses both IID and non-IID settings. In the IID scenario, each MSMS trains its model across all classes, while in non-IID setup, each MSMS is assigned a single class.


{\bf Environment.} We employed Python 3.8.19, Torch 1.11, and CUDA 10.2 as our experimental environment. We executed our FedMining approach on an NVIDIA GeForce H100 GPU with 80 GB of memory. For further details on datasets and hyperparameters, please refer to Table~\ref{tab:param_new}, which summarizes the key parameters used in our simulation with the remaining parameters consistent with those in \cite{yang2023open}.


   

{\bf Baselines.} To evaluate the security and robustness of FedMining's secure aggregation scheme, we analyzed its security guarantees, communication cost, and computational overhead. Specifically, we compared FedMining against widely used approaches based on the Paillier cryptosystem, such as Privacy-Preserving Federated Learning (PPFL)~\cite{park2022privacy}.




\subsection{Security analysis of FedMining}\label{sec:secure_eval}

FedMining is a secure FL approach to safeguard both the MSMSs' local models and their training data against unauthorized access. 
In the meantime, it achieves the {\em same security level} as other SOTA methods because of its incorporation of IPFE ~\cite{abdalla2018multi}. In the following, we explain how FedMining protects privacy against the threats discussed in \sref{Sec:Threat}:
\begin{table}[!t]
\centering
\renewcommand{\arraystretch}{1.25}
\setlength{\tabcolsep}{0.8em}
\caption{Summary of Datasets, Models, and Training Configurations.}
\label{tab:param_new}
\resizebox{\linewidth}{!}{
\begin{tabular}{lccccccc}
\toprule
\multicolumn{8}{c}{\textbf{Dataset and Model Specifications}} \\
\midrule
\textbf{Dataset} & \textbf{Size} & \textbf{Classes} & \textbf{Model} & \textbf{\# Params} & \textbf{Epochs} & \textbf{Batch} & \textbf{Weight Decay} \\
\midrule
CIFAR-10   & 60,000  & 10  & CNN      & 7{,}759{,}521  & 50  & 56 & $1\!\times\!10^{-3}$ \\
CIFAR-100  & 60,000  & 100 & CNN      & 7{,}759{,}521  & 100 & 32 & $1\!\times\!10^{-3}$ \\
DsLMF+     & 138,004 & 6   & YOLOv7   & 37{,}196{,}556 & 30  & 16 & $5\!\times\!10^{-4}$ \\
\midrule\midrule

\multicolumn{8}{c}{\textbf{Training Hyperparameters}} \\
\midrule
\textbf{Model} & \multicolumn{2}{c}{\textbf{LR Range}} & \textbf{Optimizer} & \textbf{Momentum} & \multicolumn{3}{c}{---} \\
\midrule
CNN     & \multicolumn{2}{c}{$0.001$–$0.1$} & SGD  & $0.85$  & \multicolumn{3}{c}{---} \\
YOLOv7  & \multicolumn{2}{c}{0.01}          & Adam & $0.937$ & \multicolumn{3}{c}{---} \\
\bottomrule
\end{tabular}}\vspace{-0.5cm}
\end{table}

\begin{itemize}[leftmargin=*]
    \item In scenarios where potential eavesdroppers could intercept the transmission of MSMSs' local models to the $\mathcal{AE}$, FedMining ensures that neither model parameters nor corresponding private training data can be inferred. This is because  the $\mathcal{AE}$ can only calculate the aggregated global model using the aggregation key $\mathcal{AK}_\mathcal{AE}$, without being able to compute each individual MSMS's model parameters. Please note that our design also considers the possibility of the $\mathcal{AE}$ attempting to infer any private data from the MSMSs.

    \item FedMining diverges from the conventional use of the $\mathcal{MSK}$ for all clients (MSMSs) as in traditional IPFE. Instead, FedMining has distributed secret keys, where each MSMS utilizes a distinct secret key extracted from the $\mathcal{MSK}$, which remains exclusively known to the KDC. This enhances the security of our FL framework, as intercepted encrypted model parameters of an MSMS $k$ cannot be decrypted without the corresponding unique secret key ($\mathcal{SK}_k$). Despite having distributed secret keys, the underlying encryption/decryption schemes bear resemblance to the IPFE scheme. Therefore, our approach upholds the same security level as the IPFE scheme, as demonstrated in \cite{abdalla2018multi}.


    \item FedMining is also designed to be resilient against collusion attacks between MSMSs and the $\mathcal{AE}$. Specifically, since the $\mathcal{AE}$ cannot access individual MSMSs' models, the only feasible method for it to obtain such information is through collusion with all MSMSs except the target MSMS, denoted as $k'$. In this scenario, the $\mathcal{AE}$ would perform the subtraction operation $\boldsymbol{w}^{t}-\sum_{k=1,k\neq k'}^{K}\boldsymbol{w}_k^{t} = \boldsymbol{w}_{k'}^{t}$. However, the likelihood of the $\mathcal{AE}$ successfully colluding with $K-1$ MSMSs is possible. This is because there are typically other MSMSs operated by the same vendor as the ``victim'' MSMS $k'$, and they are unlikely to collaborate with the $\mathcal{AE}$.

\end{itemize}

{\bf Comparison with SOTA.} FedMining is compared with our baseline approach PPFL \cite{park2022privacy}, where PPFL partially fulfills specific security objectives discussed above. For instance, PPFL suffers from the following security problems. First, to avoid collusion between MSMSs and $\mathcal{AE}$, PPFL needs a gateway to aggregate the participating MSMSs' model parameters and then send an encrypted version of the aggregated result to the $\mathcal{AE}$; the $\mathcal{AE}$ then needs to decrypt the result to obtain the global model (and sends it back to MSMSs). Having such a gateway is challenging because PPFL requires it to be {\em completely trusted and not collude with any entity}. Second, since the $\mathcal{AE}$ possesses the private key, it can share this key with the gateway, thereby enabling the gateway to decrypt individual MSMS models. This poses security and privacy risks to the MSMSs' local models, but is not a concern in FedMining where the $\mathcal{AE}$ only has its own key $\mathcal{SK}_{\mathcal{AE}}$. 
FedMining is more robust and ensures the MSMSs' sensitive models and data remain secure even in the presence of potential internal and external attacks mentioned in \sref{Sec:Threat}.

\subsection{Communication \& Computation Overheads of FedMining}\label{sec:privacy_comp}
\begin{table}[!t]
\caption{Comparison of FedMining's {\bf \em communication} overheads on various datasets.}
\label{train_cost}
\centering
\begin{tabular}{p{3cm}|p{2cm}}
\toprule
Model-Dataset & Size (MB) \\
\midrule
CNN using CIFAR-10 & 434.53 \\
CNN using CIFAR-100 & 434.53 \\
YOLOv7 using DsLMF+ & 2,083.01 \\
\bottomrule
\end{tabular}\vspace{-0.3cm}
\end{table}
\subsubsection{FedMining Communication Overhead Analysis.}\label{sec:comm}
The analysis of communication overhead involves measuring the combined size of models transmitted between the $\mathcal{AE}$ and participating MSMSs in each FL round. We follow the recommendation of the National Institute of Standards and Technology (NIST) by utilizing a 224-bit security-level elliptic curve for cryptographic operations. In our approach, each MSMS sends encrypted matrices containing the updated local model parameters, including weights and biases. Consequently, each element in the cyclic group $\mathbb{G}$ requires 56 bytes for representation. Therefore, FedMining will incur a total communication overhead due to encryption of $56 \times |\boldsymbol{w}_{k}^{t}|$ bytes of data in each FL round, where $|\boldsymbol{w}_{k}^{t}|$ represents the number of model parameters (not bytes). Based on the YOLOv7 model architecture (consisting of 37,196,556 parameters) applied to the DsLMF+ dataset \cite{yang2023open}, the communication overhead amounts to approximately 2,083.01~MB.


{\bf Comparison with SOTA.} PPFL~\cite{park2022privacy} yields a size of $512$ bytes for each encrypted model parameter, resulting in a total communication overhead of $512\times |\boldsymbol{w}_{k}^{t}|$ bytes of data in each FL round. Furthermore, there is a remarkable delay caused by PPFL's distributed aggregation, which requires {\em three communication rounds} with the participating MSMSs because the $\mathcal {AE}$ is unable to decrypt the aggregated parameters directly and needs three communication rounds to construct the global model securely in each FL round. This results in the total communication overhead with PPFL being $3 \times 512 \times |\boldsymbol{w}_{k}^{t}|$ = $1536 \times |\boldsymbol{w}_{k}^{t}|$ bytes. Considering our model architecture with the DsLMF+ dataset \cite{yang2023open}, this results in approximately 57,233.91~MB overheads.  In contrast, FedMining streamlines the transmission of encrypted model parameters directly to the $\mathcal{AE}$ within a single round, leading to a significant reduction in communication overhead, surpassing 27$\times$ compared to PPFL. Table~\ref{train_cost}  highlights the communication overheads of FedMining when each MSMS trains various models on different datasets, which amounts to approximately 2,083MB—substantially lower than PPFL's 57,233MB.





\subsubsection{FedMining Computation Overhead Analysis.}\label{sec:comp}
For measuring the computation overhead of our encryption scheme, we implemented it using the Python Charm library. We assessed both the encryption overhead on the MSMSs and the aggregation overhead by the $\mathcal {AE}$ in each FL round.

{\bf Comparison with SOTA.} Our measurements indicate that the computation required for each MSMS to encrypt its model parameters, $|\boldsymbol{w}_{k}^{t}|$, {\em  takes less than 3 msec using FedMining, whereas PPFL takes more than 41.6 msec, indicating a substantial $14\times$ difference}. The key factor distinguishing FedMining is its reliance on a single exponentiation operation, as opposed to the three exponentiation operations required by PPFL.
\begin{table}[!t]
\caption{ Comparison of FedMining's {\bf \em computation} overheads on various datasets.}
\label{train_cos_com}
\centering
\begin{tabular}{p{3 cm}|p{2cm}}
\toprule
Model & FLOPS (G) \\
\midrule
CNN using CIFAR-10 & 36.71 \\
CNN using CIFAR-100 & 48.36 \\
YOLOv7 using DsLMF+ & 105.11 \\
\bottomrule
\end{tabular}\vspace{-0.3cm}
\end{table}

Regarding aggregation overhead, the $\mathcal{AE}$ collects encrypted models from ${K}$ MSMSs, or fewer if some MSMSs are unavailable or unwilling to participate. The dominant operation in this process is multiplication in FedMining, while PPFL relies on modular exponentiation~\cite{park2022privacy}, which has a 56-68\% higher computational cost. Additionally, the three communication rounds required by PPFL incur additional computation. Moreover, PPFL requires client involvement in the aggregation process, where each client computes a decryption share during each FL round, resulting in additional overhead due to multiple modular exponentiation operations. Conversely, FedMining delegates the aggregation process entirely to the $\mathcal{AE}$, which might increase the overheads on the $\mathcal{AE}$ but could be manageable given its high computational capabilities.  However, implementing a similar approach to PPFL, which increases the overhead for the MSMSs, could result in elevated computation burdens for them, posing a significant issue. Indeed, by enabling the $\mathcal{AE}$ to utilize a lookup table for the pre-computation of discrete logarithms~\cite{shoup1997lower}, we can further reduce aggregation overhead. This strategy proves especially effective since $\sum_{k=1}^{K}{\alpha_{k}^t}\boldsymbol{w}_{k}^{t}$ is a small value.

Finally, Table~\ref{train_cos_com} summarizes the computation overheads of FedMining when each MSMS trains various models on different datasets. Measured in floating-point operations per second (FLOPS) for a more general assessment, the highest computation cost incurred by our approach is 105.11 GFLOPS on the DsLMF+ dataset, demonstrating its efficiency.

\subsection{FedMining Convergence Analysis}\label{sec:conv_ana}
  
{\bf Evaluating FedMining Convergence.} To evaluate FedMining's convergence speed and accuracy on the DsLMF+ dataset \cite{yang2023open}, various evaluation metrics are employed, including accuracy (ACC), precision (PC), and recall (RC). Additionally, we use average precision (AP) at two levels: 0.5 and 0.5:0.95. AP represents the area under the Precision-Recall curve, calculated as $AP=\int_{0}^{1} (PC)(RC)d(RC)$. The mean AP (mAP) provides a weighted average of AP values across all sample categories, reflecting the model's detection performance across all classes.  

In Table \ref{Fedining_eval}, we display our results of testing the final converged global model of FedMining in all metrics. As depicted, the final global model successfully achieves high performance on all considered metrics, even on mAP with 0.5:0.95 level, with an average accuracy among all classes of 83\%. These results underscore the effectiveness of the FedMining approach in achieving convergence with high accuracy across various dataset classes.

\begin{table}[!t]
\setlength{\tabcolsep}{0.4em}
\renewcommand{\arraystretch}{1}
\caption{FedMining's evaluation on DsLMF+ Dataset.} 
\label{Fedining_eval}
\centering
\begin{tabular}{p{2.95cm}|p{0.8cm}|p{0.85cm}|p{1.25cm}|p{1.85cm}}
\toprule
\textbf{Class Metric} & \textbf{PC} & \textbf{RC} & \textbf{mAP (0.5)} & \textbf{mAP (0.5:0.95)}\\
\midrule
Towline & 0.989 & 0.991 & 0.998 & 0.938 \\
\midrule
Coal miner & 0.964 & 0.976 & 0.981 & 0.872 \\
\midrule
\footnotesize Miner safety helmet & 0.951 & 0.983 & 0.974 & 0.720 \\
\midrule
Miners' behavior & 0.913 & 0.919 & 0.934 & 0.792 \\
\midrule
Large coal & 0.832 & 0.815 & 0.891 & 0.776 \\
\midrule
\scriptsize Hydraulic support plate & 0.981 & 0.937 & 0.989 & 0.884 \\
\midrule
\rowcolor{red!10} \textbf{Average} & \textbf{0.938} & \textbf{0.937} & \textbf{0.961} & \textbf{0.830} \\
\bottomrule
\end{tabular}
\end{table}
\begin{figure}[!t]
    {\subfloat[\scriptsize {mAP (0.5).}]{\includegraphics[width=0.24\textwidth]{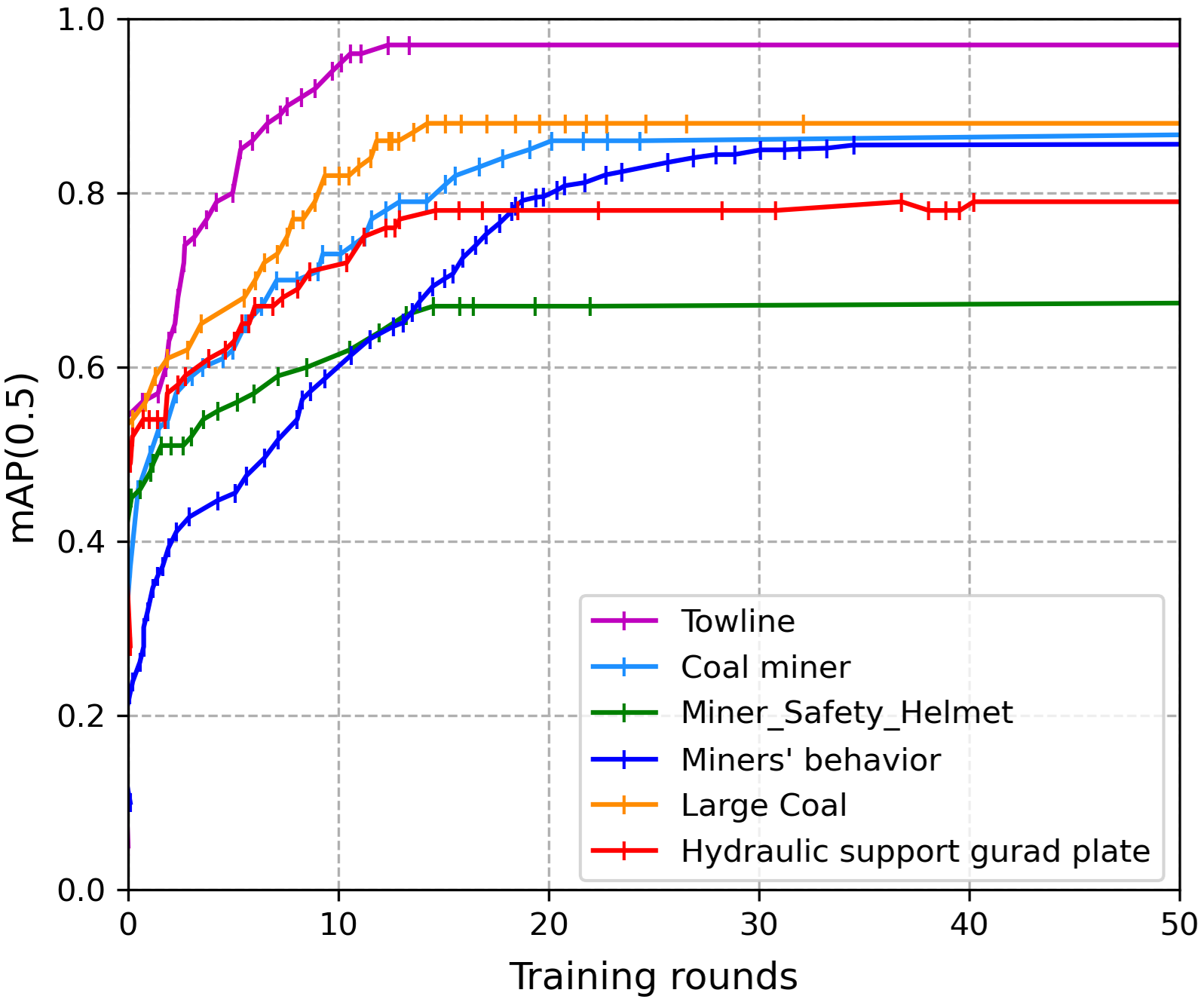}}}
    {\subfloat[\scriptsize {mAP (0.5:0.95).}]{\includegraphics[width=0.24 \textwidth]{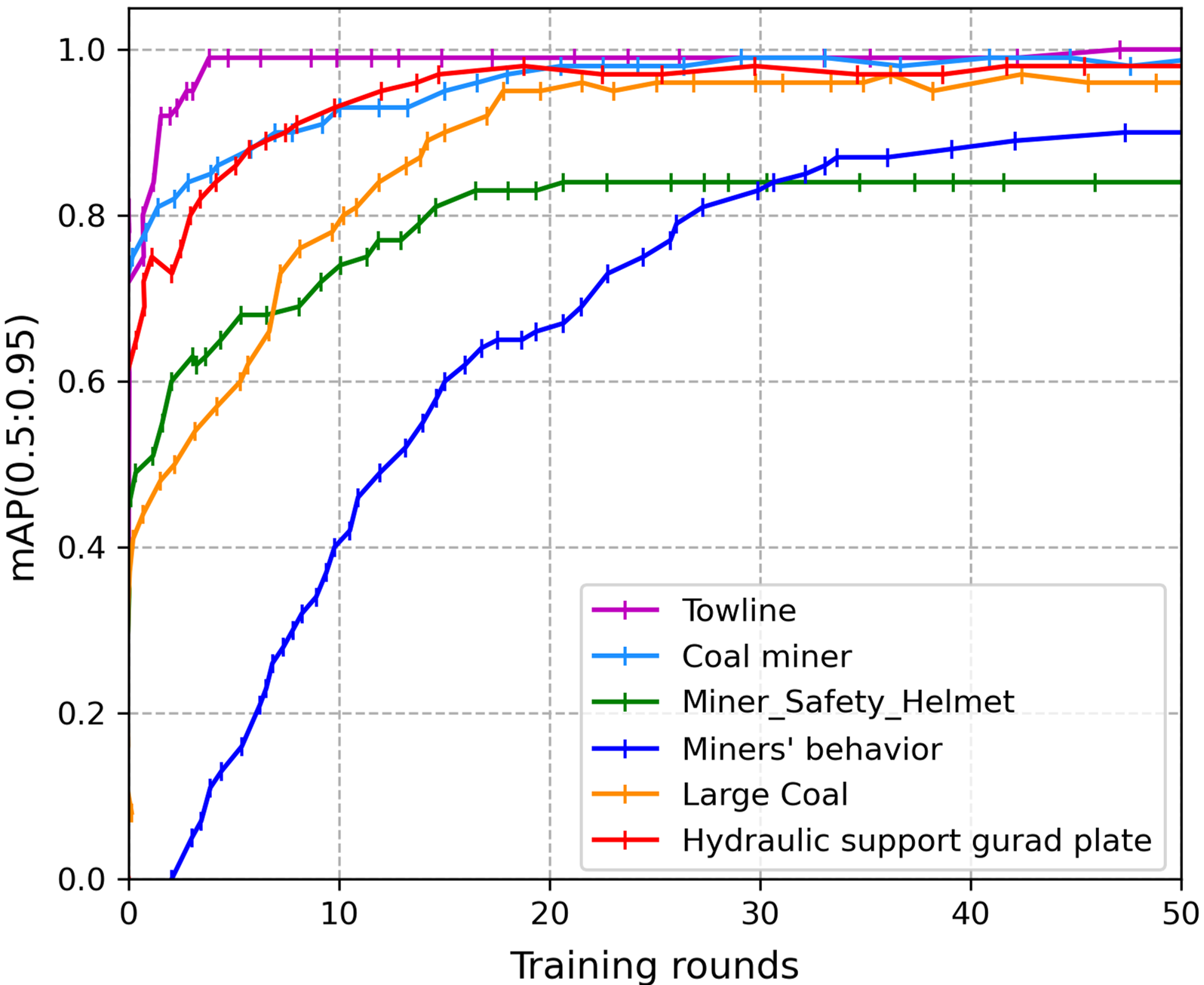}}}
\caption{Evaluating mAP changes during training across various DsLMF+'s classes. }\label{AP_fig}
\end{figure}

Additionally, in \fref{AP_fig}, we present the convergence results of the global model using mAP at two levels for all the classes across multiple communication/training rounds. The findings indicate that FedMining's global model begins to converge only after 20 rounds with an average AP surpassing 80\% at the mAP 0.5 level. This demonstrates the rapid convergence of our approach while safeguarding the privacy of MSMSs' local models and their associated data.
Furthermore, when considering the mAP levels from 0.5 to 0.95, we observe that FedMining's global model also starts to converge on all classes, but after 30 rounds, maintaining high accuracy comparable to the mAP 0.5 level.

{\bf Evaluating the Balancing Aggregation Scheme.} To investigate if the global model exhibits biases towards specific classes, we conduct a detailed analysis of its performance on individual classes. This evaluation aims to assess its effectiveness in accurately detecting all classes, particularly in a non-IID setting where each MSMS was trained on a single class from the DsLMF+ dataset.

\begin{figure} 
\centering
\includegraphics[width=0.9\linewidth]{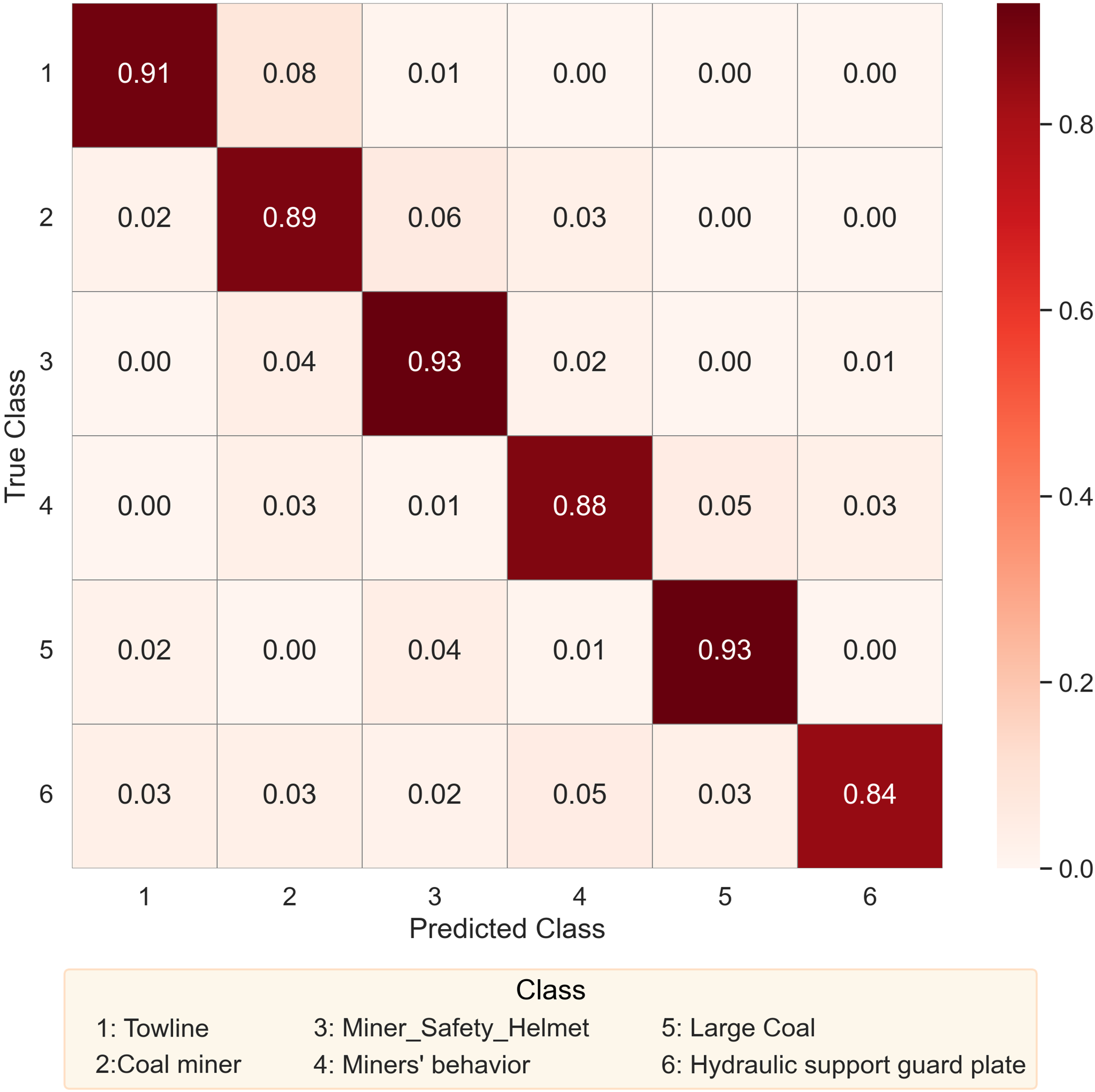}
\caption{Confusion matrix depicting the correspondence between predicted outputs of the global model and ground-truth class labels.} \label{conf_mtrx}
\end{figure}

In \fref{conf_mtrx}, we display the confusion matrix illustrating the predicted accuracy versus the ground truth across the six classes. Notably, FedMining consistently achieves an accuracy rate of 84\% or higher for all classes, demonstrating its effectiveness in diverse class detection scenarios. These results highlight the robustness and fairness of FedMining, especially in challenging training environments with non-IID settings, where the global model successfully predicts all classes with high accuracy.


\begin{figure*}[!t] 
 
\centering
\begin{subfigure}{\linewidth}
    \includegraphics[width=\linewidth]{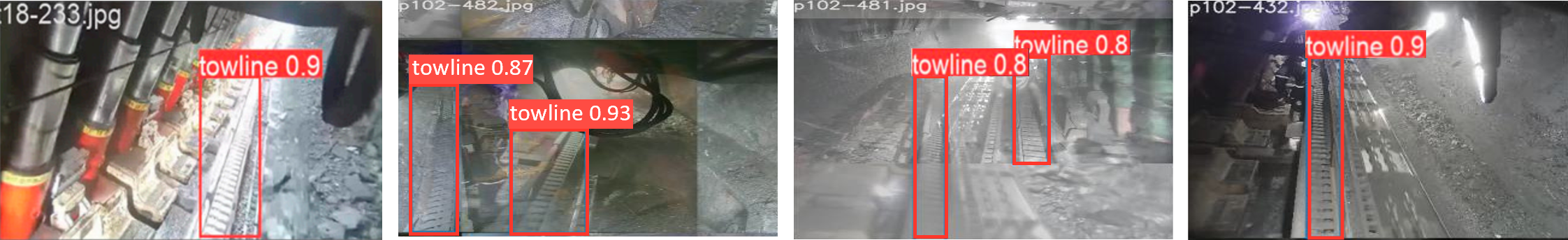}
    \caption{Towline.}
\end{subfigure}
\begin{subfigure}{\linewidth}
    \includegraphics[width=\linewidth]{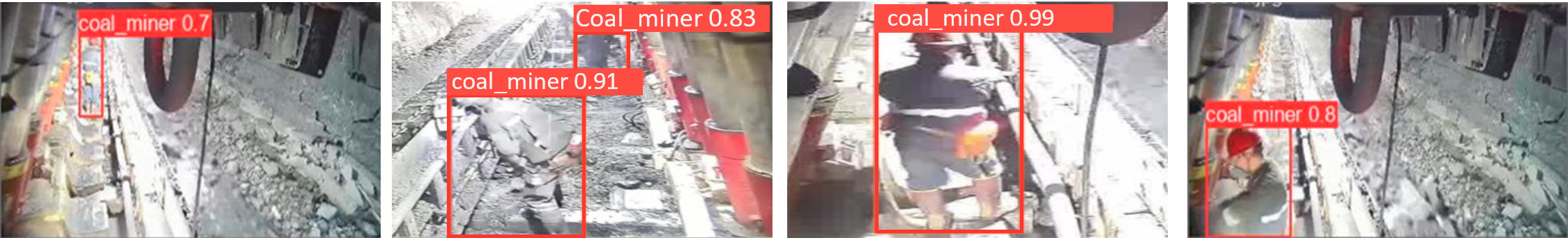}
    \caption{Coal miners.}
\end{subfigure}
\begin{subfigure}{\linewidth}
    \includegraphics[width=\linewidth]{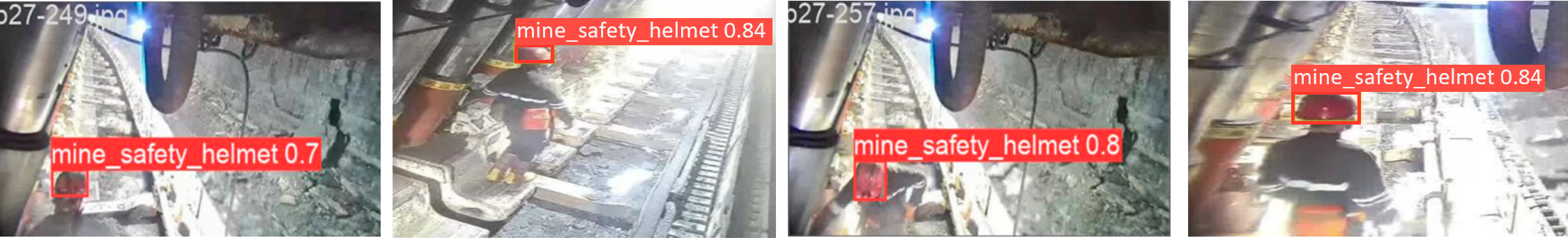}
    \caption{Miner Safety helmet.}
\end{subfigure}
\begin{subfigure}{\linewidth}
    \includegraphics[width=\linewidth]{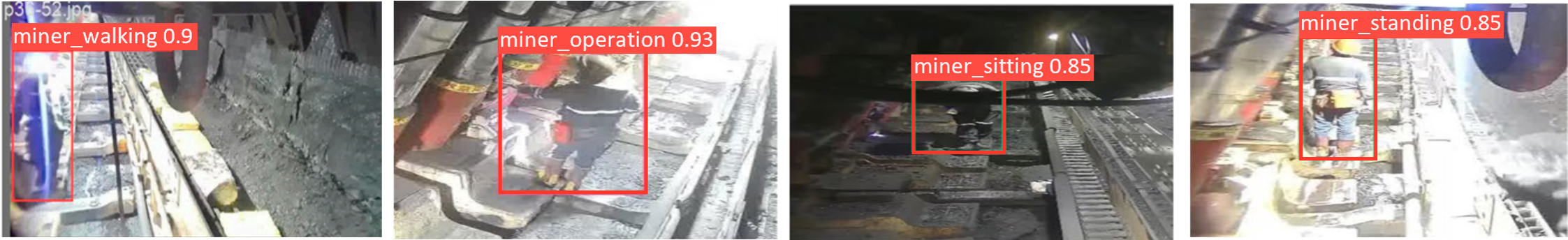}
    \caption{Miners' behaviors.}
\end{subfigure}
\begin{subfigure}{\linewidth}
    \includegraphics[width=\linewidth]{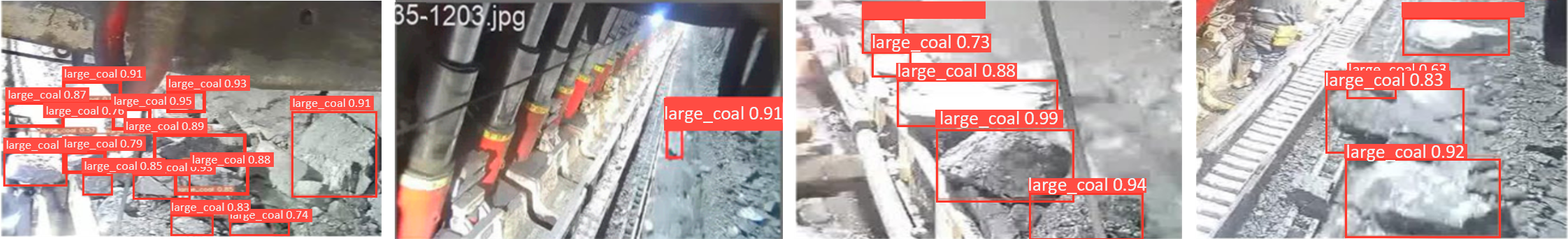}
    \caption{Large Coal.}
\end{subfigure}
\begin{subfigure}{\linewidth}
    \includegraphics[width=\linewidth]{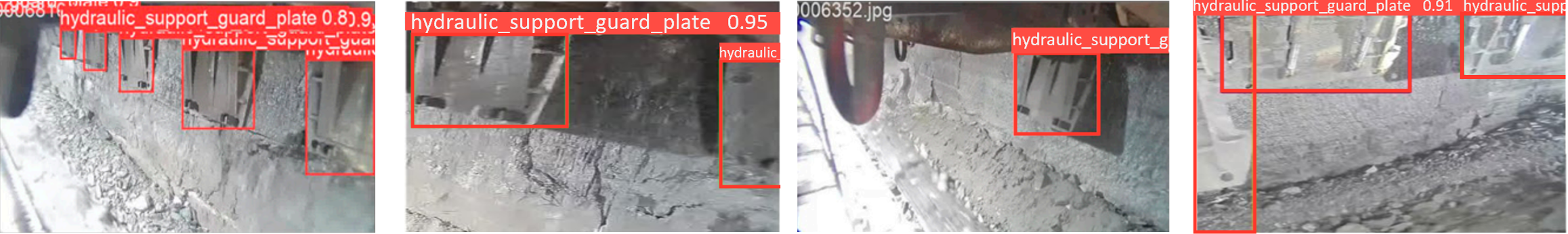}
    \caption{Hydraulic support guard plate.}
\end{subfigure}
\caption{Samples from the DsLMF+ dataset used to test the prediction accuracy of the global model.}
\label{fig:precision}
 
\end{figure*}
Furthermore, in \fref{fig:precision}, we present the detection results for six objects, offering insight into the object detection accuracy attained by the FedMining approach. The depicted samples demonstrate FedMining's consistent ability to detect all classes with high accuracy levels, effectively identifying objects with comparable precision. This underscores the effectiveness of our proposed balancing aggregation scheme, which enables FedMining to yield {\em a global model capable of detecting various objects, including those for which individual MSMSs were not trained}.

\section{Conclusion}

In this paper, we proposed FedMining, a privacy-preserving FL approach tailored for detecting hazards to miner safety in underground mining environments. FedMining ensures the privacy of MSMSs' local models and associated private data while achieving rapid convergence and high accuracy across various classes. It addresses security challenges by protecting individual MSMSs' models from potential adversaries attempting to breach data privacy through model inversion attacks or membership inference. It innovates through two key components: (i) a {\em secure aggregation scheme} that encrypts MSMSs' model parameters during FL training using an efficient IPFE approach, ensuring that neither the FL server nor any eavesdroppers can access MSMSs' raw data, even in insecure communication channels; (ii) a {\em balancing aggregation scheme} to handle non-IID data collected by individual MSMSs and their availability, enabling the generation of a global model capable of accurately classifying diverse hazards classes. Our experiments on real mining datasets demonstrate FedMining's effectiveness in preserving data privacy while achieving an average convergence accuracy of 93\%. Furthermore, FedMining incurs low communication and computation overheads compared to SOTA methods, highlighting the lightweight nature of our encryption and aggregation scheme.

\bibliographystyle{IEEEtran}
{\small
\bibliography{_references.bib}}

\end{document}